\documentclass[aps,pre,10pt,groupedaddress,amsfonts,amssymb,amsmath,showpacs,showkeys,a4paper,floatfix,notitlepage]{revtex4-1}

\usepackage[english]{babel}
\usepackage[T1]{fontenc}
\usepackage{lmodern}
\usepackage[utf8]{inputenc}
\usepackage[protrusion=true,expansion=true]{microtype}
\usepackage{amsmath,amsfonts,amsthm}
\usepackage{ bbold }
\usepackage[pdftex]{graphicx}
\usepackage[outdir=./]{epstopdf}
\usepackage{epstopdf}
\usepackage{bm}
\usepackage[usenames,dvipsnames,svgnames,table]{xcolor}
\usepackage{csquotes}
\usepackage{booktabs} 
\usepackage{array} 
\usepackage{geometry}
\usepackage{MnSymbol}
\usepackage{pgf}
\usepackage{import}
\usepackage{caption}
\usepackage{subcaption}
\usepackage{wrapfig} 
\usepackage[normalem]{ulem} 
\usepackage{comment} 

\usepackage{lineno,hyperref}

\usepackage{graphicx}

\usepackage{epstopdf, epsfig}
\usepackage{caption}
\usepackage{subcaption}
\usepackage[usenames,dvipsnames,svgnames,table]{xcolor}
\newcommand{\degree}{\ensuremath{^\circ}}
\usepackage{float}

\usepackage{amssymb,amsmath}
\usepackage{bm}
\usepackage{natbib}

\usepackage{bm}
\usepackage{multirow}
\usepackage{mathrsfs}

\usepackage{multirow}
\newif\ifchanges
\changestrue

\begin{document}

\title{Entropic Lattice Boltzmann Method for Moving and Deforming Geometries in Three Dimensions}


\author{B. Dorschner}
\affiliation{Aerothermochemistry and Combustion Systems Lab,Department of Mechanical and Process Engineering, ETH Zurich, CH-8092 Zurich, Switzerland}

\author{Shyam S. Chikatamarla}
\affiliation{Aerothermochemistry and Combustion Systems Lab,Department of Mechanical and Process Engineering, ETH Zurich, CH-8092 Zurich, Switzerland}

\author{Ilya Karlin}
\email{karlin@lav.mavt.ethz.ch}
\affiliation{Aerothermochemistry and Combustion Systems Lab,Department of Mechanical and Process Engineering, ETH Zurich, CH-8092 Zurich, Switzerland}

\date{\today}

\begin{abstract} 
Entropic lattice Boltzmann methods have been developed to alleviate intrinsic stability issues of lattice Boltzmann models for under-resolved simulations.
Its reliability in combination with moving objects was established for various 
laminar benchmark flows in two dimensions in our previous work \citet{Dorschner2015} as well as for three dimensional one-way coupled 
simulations of engine-type geometries in \citet{Dorschner2016} for flat moving walls.
The present contribution aims to fully exploit the  advantages of entropic lattice Boltzmann models in terms of stability and accuracy and extends 
the methodology to three-dimensional cases including two-way coupling between fluid and structure, turbulence and deformable meshes.
To cover this wide range of applications, the classical benchmark of a sedimenting sphere is chosen first to validate the general two-way coupling algorithm. 
Increasing the complexity, we subsequently consider the simulation of a plunging SD7003 airfoil at a Reynolds number of ${\rm Re}=40000$ and finally, to access the model's performance for deforming meshes, we conduct a two-way coupled simulation of a self-propelled anguilliform swimmer. 
These simulations confirm the viability of the new fluid-structure interaction lattice Boltzmann algorithm to simulate flows of engineering relevance.
\end{abstract}

\maketitle

\section{Introduction}
Flows induced by complex and moving geometries are of paramount interest in many fields ranging from 
industrial applications for the optimization of internal combustion engines, turbines, etc., to flows in the field of 
bio-fluidmechanics in order to enhance our basic understanding of the propulsion mechanisms of animals, the flow through the cardiovascular system 
and many more (see, e.g., \citet{Howell2010,Schmitt2014a,Kern2006, DeTullio2009,Randles2013}).
Besides the valuable insight from experimental studies, light may be shed on these complex mechanisms 
through numerical simulations.
Thus, accuracy, robustness and efficiency of the numerical treatment is crucial.\\
In this context, the lattice Boltzmann method (LBM) gains increasing attention due to its accuracy and reliability in various regimes including turbulence, thermal flows, compressible flows, micro flows, porous media and multiphase flows among many others (see \citet{Ansumali2006c,Bosch2015,Frapolli2014,Frapolli2015,Mazloomi2015,Mendoza2010}).
The LBM originates from kinetic theory and describes the flow in terms of discretized particle distribution functions (populations) $f_i(\bm x, t)$ associated with a set of discrete velocities $\bm c_i, i=1,\dots,Q$, where the dynamics of the populations are constructed to recover the desired macroscopic equations in the hydrodynamic limit. 
Efficiency is achieved by organizing the discrete velocities into a regular lattice, resulting in a stream-and-collide algorithm with exact propagation and spatially local non-linearity implemented through the collision operator. 
{The first realization, which made a practical connection of LBM to fluid dynamics was the well known lattice Bhatnagar-Gross-Krook (LBGK) model. The LBGK model remains popular  due to its simplicity and numerical efficiency.}
Despite these attractive properties, the LBM was not competitive for long due to the lack of numerical stability and accurate boundary conditions in 
the simulation of high Reynolds number flows. 
This promoted the development of various models intended to resolve this issue, including explicit turbulence models in LBM formulation, 
multiple-relaxation time models (MRT) and others (see, e.g., \citet{Karlin1999,Lallemand2000,Chen2003,Latt2006,Geier2006}).
Most notably, the entropic lattice Boltzmann method (\citet{Karlin1999}), which, by introduction of a discrete entropy function $H$ as the determining 
factor of the relaxation, features non-linear stability. 
In contrast to LBGK, the relaxation parameter is chosen adaptively at each point in space and time to locally ensure the discrete-time H-theorem. Consequently, an effective viscosity is prescribed by the second law of thermodynamics, which may both smoothen or enhance 
the flow field.
Recently, the entropic concept was extended to multi-relaxation time models (KBC models), 
where the additional freedom of multiple relaxation parameters is used to keep the effective viscosity at its nominal value while retaining stability. 
Excellent performance of KBC models in terms of stability and accuracy 
for under-resolved simulations have been reported in the works of \citet{Karlin2014,Bosch2015,Dorschner2016}.
Due to its advantages, all simulations in the remainder of the paper are conducted using the KBC model.\\
In order to establish these entropy-based lattice Boltzmann models as a robust and predictive tool for complex engineering problems, 
the topic of boundary conditions needs to be addressed as its
implementation is crucial for accuracy and stability of the entire simulation.
Various proposals for flat, curved and moving boundaries can be found in the literature.
Most popular for its simplicity is the bounce-back boundary condition, where the geometry is represented using a staircase approximation.
For a more accurate representation of the geometry, common approaches use interpolation or extrapolation of the populations to account for the curvature of the object.
However, spurious shocks for turbulence simulation at the boundary limit their usage to resolved and moderate Reynolds number flows. 
An interesting realization of the bounce-back can be found for a recent, so-called crystallographic LBM, in\cite{namburi2016crystallographic}.
Another common approach for complex moving objects is the immersed boundary method (IBM), which was originally proposed by \citet{Peskin1977} for the simulation 
of blood flow. 
Notable in the LBM realm is the IBM procedure as introduced by \citet{Suzuki2011}, which resolved the common issue of streamlines penetrating the object and more 
accurately satisfies the no-slip boundary condition.\\
In our recent contribution \citet{Dorschner2015}, we aimed to resolve these issues by employing Grad's approximation to impose the desired boundary conditions on
the slow varying moments rather than the 
fluctuating population directly. 
In \citet{Dorschner2015} excellent stability properties and second-order accuracy were reported.
Further, this boundary condition was validated for various laminar benchmark simulations with stationary and moving walls in two-dimensions.
In this paper we extent these models for moving objects in three dimension with a successively increasing level of complexity.
First, we consider the classical two-way coupled benchmark simulation of a falling sphere under gravity to validate the boundary condition and the 
two-way coupling algorithm.
The next level of complexity is a simulation of a plunging airfoil at ${\rm Re}=40000$ in the transitional regime, where we report a detailed comparison 
with experimental and numerical studies of \citet{McGowan2008_AIAA} and \citet{Visbal2009}, respectively.
Finally, the performance for deformable meshes is accessed on the example of a self-propelled anguilliform swimmer. \\
The paper is organized as follows: 
In section \ref{sec:KBC}, the entropic multi-relaxation time model is reviewed followed by a brief exposure of the boundary conditions in section 
\ref{sec:bcs}.
Subsequently, in section \ref{sec:results}, we present our results for the simulation of the sedimenting sphere, the plunging airfoil and the 
angulliform swimmer.
Finally, concluding remarks are given in section \ref{sec:conclusion}.

\section{Entropic multi-relaxation time lattice Boltzmann model - KBC model}
\label{sec:KBC}
The population $f_i(\bm x , t )$ is evolved in accordance with the discrete kinetic equation
\begin{equation}
	f_i(\boldsymbol{x+c_i}, t+1)=f_i^{\prime} = (1-\beta)f_i(\boldsymbol{x},t) + \beta f_i^{\text{mirr}}(\boldsymbol{x},t),
	\label{eq:kineticEq}
\end{equation}
where the streaming step is prescribed by the left-hand side and the post-collision state $f^\prime_i$ is given 
on the right-hand side by a convex-linear combination of $f_i(\bm{x},t)$ and the mirror state 
$f_i^{\text{mirr}}(\bm{x},t)$.
Various lattice Boltzmann schemes differ in the way the mirror state is constructed.
The KBC model as employed in this paper represents the population $f_i$ in its natural moment basis 
and partitions it into three parts as
\begin{equation}
	f_i=k_i+s_i+h_i,
\end{equation}
where the kinematic part $k_i$ depends only on the locally conserved moments.
The shear part is denoted by $s_i$ and includes the deviatoric stress tensor. All remaining higher-order moments not included 
in $k_i$ or $s_i$ are lumped into $h_i$.
The mirror state,
in Eq.~(\ref{eq:kineticEq}), can now be expressed as
\begin{equation}
	f_i^{\text{mirr}} = k_i + \left( 2 s_i^{\rm eq} -s_i \right) + \left( \left(1 - \gamma\right)  h_i + \gamma h_i^{\rm eq}\right),
	\label{eq:kbc_relax}
\end{equation}
where $s_i^{\rm eq}$ and $h_i^{\rm eq}$ denote $s_i$ and $h_i$ evaluated at equilibrium
and the parameter $\gamma$
quantifies the relaxation rate of the higher-order moments.

In the athermal case, the equilibrium populations $f_i^{{\rm eq}}$ are obtained by minimization of the discrete entropy function $H$ subject to local mass and momentum 
conservation, 
\begin{equation}
	\text{min} \left\lbrace  H(f) = \sum_i f_i \ln \left( \frac{f_i}{W_i} \right) \right\rbrace, 
	\quad \text{s.t.} \sum_i \lbrace 1, \boldsymbol{c_i} \rbrace f_i = \lbrace \rho, \boldsymbol{j} \rbrace,
	\label{eq:feq_min}
\end{equation}
where the lattice-specific weights are denoted by $W_i$.
In this paper the standard three-dimensional $D3Q27$ lattice with a speed of sound of $c_s=1/\sqrt{3}$ is used for all computations, 
where $D$ and $Q$ are the dimensionality and the number of discrete velocities, respectively.
The exact solution to the minimization problem can be found in \citet{Ansumali2003a} (see also \citet{Bosch2015} for a detailed discussion ).

The Chapamann-Enskog analysis applied to this system yields the Navier-Stokes equations with the shear viscosity $\nu$ and the bulk viscosity 
$\xi$ as
\begin{equation}
	\nu=c_s^2\left( \frac{1}{2 \beta} - \frac{1}{2} \right), \quad 	\xi= c_s^2 \left( \frac{1}{\gamma \beta} - \frac{1}{2} \right).
\end{equation} 
As for any MRT model trying to increase stability, the KBC model exploits the fact that the higher-order moments may be relaxed independently without 
affecting the hydrodynamic limit.
However, unlike in the conventional MRT approach, the relaxation parameter $\gamma$ in the KBC model is not a constant tuning  parameter  
but is rather computed locally in every time step and at every grid point by minimizing the discrete entropy function 
in the post-collision state $f_i^{\prime}$.
The minimization results in the following equation for the stabilizer $\gamma$:
\begin{equation}
	\sum_i \Delta h_i \ln \left[ 1+ \frac{ (1-\beta \gamma) \Delta h_i - (2 \beta -1 ) \Delta s_i }{ f_i^{ \text{eq} } } \right] =0,
	\label{eq:gamma_min}
\end{equation}
where $\Delta s_i = s_i - s_i^{\text{eq}}$ and $\Delta h_i=h_i-h_i^{\text{eq}}$.
Moreover, an analytic approximation for the stabilizer $\gamma$ may be obtained by expansion of Eq.~(\ref{eq:gamma_min}) to the first non-vanishing order 
of $\Delta s_i/f_i^{\rm eq}$ and $\Delta h_i/f_i^{\rm eq}$ as
\begin{equation}
	\gamma = \frac{1}{\beta} - \left( 2 - \frac{1}{\beta}\right) \frac{\left< \Delta s | \Delta h\right>}{\left< \Delta h | \Delta h\right>  },
\end{equation}
where the entropic scalar product $\left< X | Y \right> = \sum_i (X_i Y_i / f_i^{\text{eq}})$ is introduced to ease notation.
The accuracy of this approximation has proven to be sufficient for all simulations reported in this paper.
For an in-depth discussion on the entropic multi-relaxation time models, the reader is referred to \citet{Bosch2015}.

\section{Wall-Boundary conditions}
\label{sec:bcs}
We here aim at extending the capabilities of KBC models to an accurate implementation of 
moving and deformable objects. 
Thus, the boundary conditions as detailed in \citet{Dorschner2015} are briefly summarized.

In order to complete the streaming step in the LB algorithm, the set of populations $\bar{D}$ advected from a solid node ${\bm x}_s$ to a fluid boundary node ${\bm x}_b$ are unknown 
and require to be specified in accordance with the desired boundary conditions.
In \citet{Dorschner2015} we proposed to approximate these "missing" populations by an analog of Grad's distribution function. 
This results in a parametrization of the distribution in terms of relevant moments, where not only locally conserved but also other pertinent moments 
may be taken into account.
It is worth noting that this is in line with the notion of minimal entropy as used in bulk for the KBC model as the Grad distribution may analogously
be derived using minimum entropy or quasi-equilibrium considerations as discussed in \citet{Gorban2010}.
In the athermal case as considered in this paper, it suffices to include the locally conserved quantities along with the pressure tensor $\boldsymbol{\Pi}$.
Explicitly, the Grad distribution with those contributions reads
\begin{equation}
	f_i^{\ast}(\rho, \bm{u}, \bm{\Pi})= W_i \left[ \rho + \frac{\rho }{c_s^2} \bm{c_i}\cdot \bm{u} + 
															  \frac{1}{2 c_s^4} \left( \bm{\Pi} - \rho c_s^2 \bm{I} \right) :
															  \left( \bm{c_{i}} \otimes \bm{c_{i}} - c_s^2 \bm{I}\right)
												    \right],
\label{eq:grad}
\end{equation}
where the pressure tensor is approximated as 
\begin{equation}
\bm{\Pi}=\bm{\Pi}^{\text{eq}}+\bm{\Pi}^{\text{neq}},
\end{equation}
with 

\begin{align}
\label{eq:pressure}
\bm{\Pi}^{\text{eq}}  &= \rho c_s^2 \bm{I}+ \rho \bm{u} \otimes \bm{u}, \\
\bm{\Pi}^{\text{neq}} &= -\frac{\rho c_s^2}{2 \beta}   \left(   \nabla \bm{u} +   \nabla \bm{u}^\dag  \right).
\label{eq:pressure2}
\end{align}
The full specification of Grad's distribution requires the density $\rho$, the velocity $\bm u $ and the pressure tensor to be prescribed.
For this purpose, the concept of target values is introduced.
The momentum exerted by the object is accounted for by specifying an appropriate target velocity $\bm{u}_{{\rm tgt}}$ at $\bm{x}_b$, which may be obtained by an interpolation involving the wall velocity \mbox{$\bm{u}_{w,i}=\bm{u}(\bm{x}_{w,i},t)$} at the intersection point $\bm{x}_{w,i}$ with the object along the velocity vector $\bm c_i$ and the velocities \mbox{$\bm{u}_{f,i}=\bm{u}(\bm{x}_{f,i}, t)$} at the adjacent fluid nodes $\bm{x}_{f,i}=\bm{x}_b+\bm{c}_i \delta_t$ for $i \in \bar{D}$. 
Using an averaged linear interpolation for the target velocity yields
\begin{equation}
\bm{u}_{{\rm tgt}}= \frac{1}{n_{\bar{D}}} \sum_{i \in \bar{D}}\frac{q_i \bm{u}_{f,i} + \bm{u}_{w,i} }{1+q_i} ,
\end{equation}
where $n_{\bar{D}}$ is the number of unknown populations and $q_i=\| \bm{x}_b-\bm{x}_{w,i} \| / \| \bm{c}_i \|$.
The target density on the other hand has two contributions corresponding to the static and the dynamic part as
\begin{equation}
	\rho_{{\rm tgt}}=\rho_{{\rm stat}} +\rho_{{\rm dyn}}	
\end{equation}
with
\begin{align}
\rho_{{\rm stat}} &=\sum_{i \in \bar{D}} f_i^{{\rm{bb}}} + \sum_{i \notin \bar{D}} f_i , \\
\rho_{{\rm dyn}} &=\sum_{i \in \bar{D}} 6 W_i \rho_0 \bm{c}_i \cdot \bm{u}_{w,i}  ,
\end{align} 
where the static part $\rho_{{\rm stat}}$ is the implied density if one were to use the bounce-back boundary condition to ensure no mass flux through the boundary. The reflected population $f_i^{{\rm{bb}}}$ is defined as $f_i^{{\rm{bb}}}=\tilde{f}_i$, where $\tilde{f}_i$ is associated with the velocity vector $\tilde{c}_i=-c_i$.
The dynamic part $\rho_{{\rm dyn}}$ accounts for the density alteration caused by the mass displacement by the moving body and may be derived by
introducing a forcing term $F_i$, which is necessary for the displacement. 
The mass and momentum conservation	$\sum_{i \in \bar{D}} F_i=0$ and $\sum_{i \in \bar{D}} \bm{c}_i  F_i=\rho \bm{u}_w$
directly lead to 
\begin{equation}
	F_i=6 W_i \rho \bm{c}_i \cdot \bm{u}_w
\end{equation}
for the $D3Q27$-lattice, where the summation over all unknown populations in $\bar{D}$ yields the implied change in density for moving objects.
Finally, the pressure tensor is prescribed by computing $\nabla \bm{u}$ using a finite difference scheme and the velocity values from the previous time step
and evaluating Eqs.~(\ref{eq:pressure}-\ref{eq:pressure2}).
Another aspect to be considered for moving objects is the reinitialization or refill of the lattice sites, which are uncovered as the 
objects passes by. 
For such nodes, we again employ the Grad distribution as given in Eq.~(\ref{eq:grad}) with the wall velocity and a local density average.
The pressure tensor is evaluated in the same manner as for the boundary conditions.

These boundary conditions constitute the action of the object onto the fluid.
On the other hand, for two-way coupled simulations, the feedback exerted from the fluid onto the object is
accounted for by solving Newton's equations of motion and will be detailed in the corresponding section below.


\section{Results and discussion}
\label{sec:results}

\subsection{Sedimenting sphere}

\label{sec:sphere}
\begin{table}[!b]
\begin{center}
\caption{\label{tab:table4}%
\label{tab:sphere}
Flow past a sphere at ${\rm Re}=100$.}
\begin{tabular}{lcc}
Contribution						&	$C_d$					&	$L/D$	\\
\hline
\citet{Johnson1999} 			&  $1.1$					& 	$0.88$		\\	
\citet{Eitel-Amor2013}			&  $1.098$				& 	$0.87$		\\	
present 							&  $1.1$					& 	$0.86$		\\	
\end{tabular}
\end{center}
\end{table} 

\begin{figure}
\centering
\includegraphics[width=0.7\textwidth]{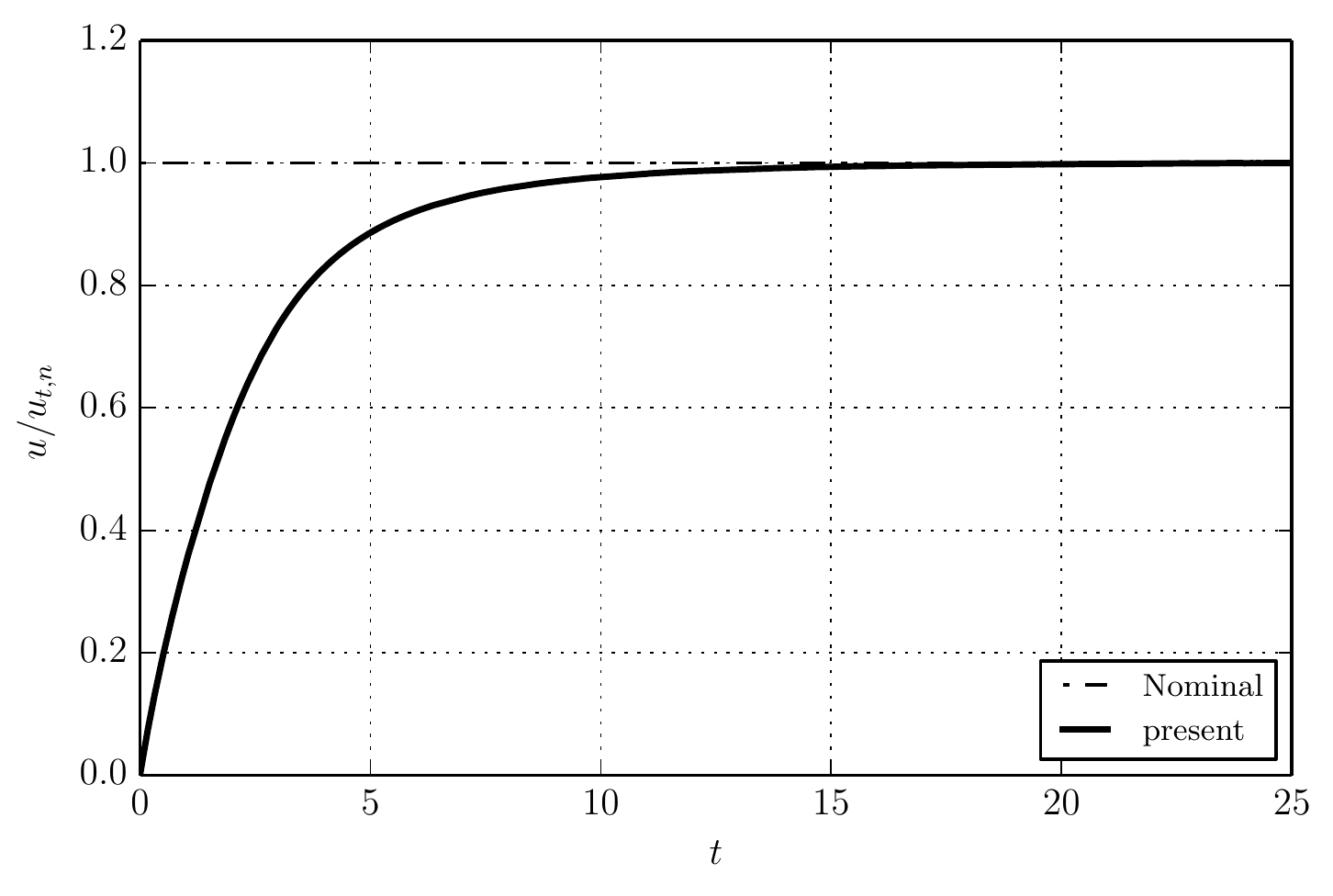}
\caption{Temporal velocity evolution of a sedimenting sphere.}
\label{fig:settlingVel_sphere}
\end{figure}

As a first step to validate the proposed two-way coupled KBC algorithm, we consider the classical benchmark of a settling particle under gravity.
For this purpose, we conduct two simulations. In the first simulation, we keep the sphere stationary and impose a mean flow in order to validate and compute the drag coefficient and the recirculation length for a Reynolds number of ${\rm Re}=u_\infty D_s/\nu=100$.
The sphere is resolved with $D_s=30$ grid points and the characteristic velocity $u_\infty=0.01$ is given by the mean flow velocity.
After the initial transient, the drag coefficient and the recirculation length are measured and the results are listed in Table (\ref{tab:sphere}) along with literature values.
With a good agreement to all reference data, we perform a second simulation for which there is no mean flow velocity but instead the sphere is settling under gravity.
When released in a quiescent fluid with density $\rho_f$, the sphere with density $\rho_s$ accelerates towards its terminal settling velocity $u_t$ for which the gravitational force $F_g=\pi D_s^3 \rho_s \text{g}/6$ is balanced by the buoyancy force $F_b=\pi D_s^3 \rho_f \text{g}/6$ and the drag force $F_d=\rho_f u^2 \pi D_s^2  C_d/8$.
Establishing the force balance relates the density ratio to the terminal velocity as 
\begin{equation}
\frac{\rho_s}{\rho_f}=1+ \frac{3 u_t^2 C_d}{4D_s\text{g}},
\label{eq:settlingSphere_forceBalance}
\end{equation}
where $C_d$ is the drag coefficient. In this two-way coupled simulation, the feedback from the fluid onto the sphere is prescribed by 
Newton's equations for the particle motion as 
\begin{align}
	\dot{\bm{x}}_s&=\bm{u}_s, \\
	\ddot{\bm{x}}_s&=\frac{\bm{F}}{m_s} + (1- \frac{\rho_f}{\rho_s}) \bm{\mathbf{g}}  , \\
\end{align}
which is solved by an Euler integration and the force $\bm F$ is computed via the Galilean invariant momentum exchange method (see \citet{Wen2015}). 
The evolution of the settling velocity is shown in Fig.~(\ref{fig:settlingVel_sphere}) for an imposed drag coefficient from the stationary simulation and a nominal settling velocity of $u_{t,n}=0.01$.
After the initial acceleration, a terminal velocity of $u_t = 0.0100033$ is reached at $t=t_{lb} u_{t,n}/D=25$ non-dimensional time units, which corresponds to less than $0.033\%$ error. 
Thus, Galilean invariance is established between the stationary and the moving case and therefore validates the basic two-way coupling algorithm.

\subsection{Plunging Airfoil}
\label{sec:airfoil}

\begin{figure}[!t]
\centering
\includegraphics[width=0.7\textwidth]{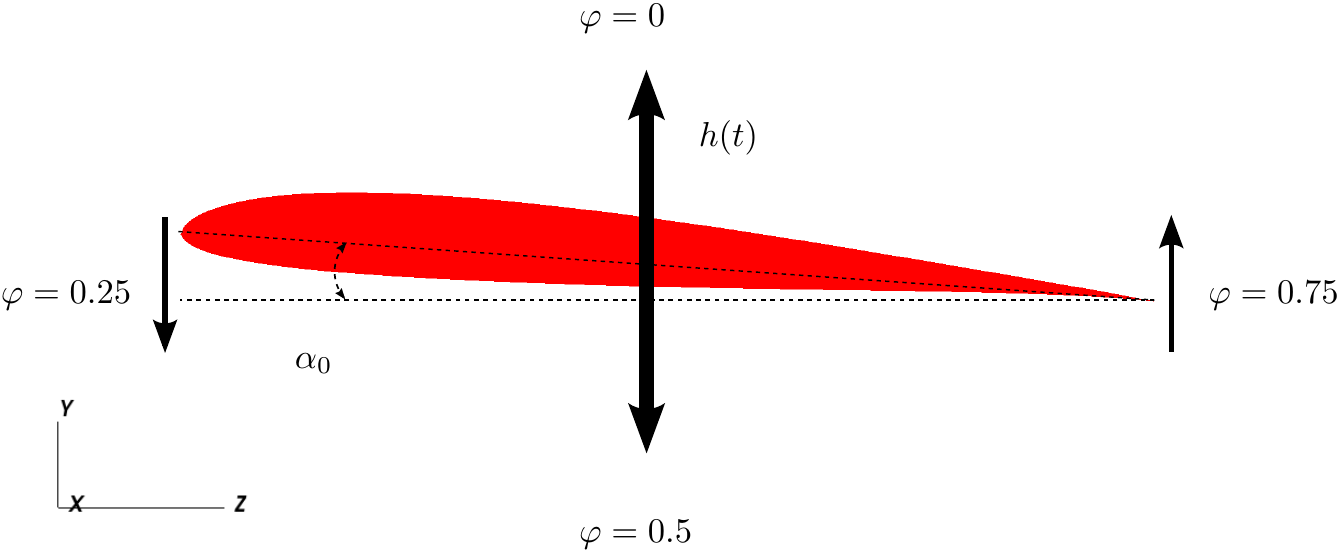}
\caption{Schematic of a plunging airfoil.}
\label{fig:schematic_airfoil}
\end{figure}

In this section, we leave the laminar flow regime and focus on the transitional flow past a plunging airfoil at a Reynolds number ${\rm Re}=40000$.
Motivated to deepen our understanding of the complex physics relevant to small fliers, small unmanned air vehicles, micro air vehicles and alike, 
this setup was recently investigated experimentally and numerically in the works of \citet{McGowan2008_AIAA,Visbal2009,Ou2011a}. 
Analogous  to small fliers, the flow is mainly characterized by the formation of dynamic-stall vortices on the leading edge due to the 
large induced angle of attack.
The transitional flow regime is particularly demanding for turbulence models as a high-Reynolds number analysis may not be valid 
in the presence of both laminar and turbulent flow. 
This gives us the opportunity to test both the KBC model and the implementation of moving boundaries to its full extent.
Notable is the recent study from \citet{Visbal2009} using an implicit Large-Eddy simulation (ILES, see, e.g., \citet{Margolin2002, Margolin2006} for details and the rationale of ILES) with high-order compact schemes for the spatial derivatives needed to capture
the transitional processes accurately and a Pade-type low-pass filter to gain stability.

In particular and as shown in Fig.~(\ref{fig:schematic_airfoil}), the flow past a plunging SD7003 airfoil with a static angle of attack $\alpha_0=4 \degree$ is considered in this contribution.
The airfoil is resolved by $L=400$ lattice points using two levels of block-refinement with a refinement ratio of two near the airfoil and a domain of 
$\left[ 10L \times 5L \times 0.2c \right]$ in the streamwise, lateral and spanwise direction, respectively. The grid refinement technique used within this context is detailed and validated in \citet{Dorschner2016a}.
Same as in \citet{Visbal2009}, a sinusoidal plunging motion of the airfoil is prescribed with a non-dimensional plunging amplitude of $h_0=h_D/c =0.05$ and a reduced frequency of $k=\pi f c /u_\infty$=3.93 as
\begin{equation}
	h(t)= h_0 \sin(2kF(t))
\end{equation}
where 
\begin{equation}
	F(t)=1-e^{at}, \quad a=4.6/2
\end{equation}
is an initial delay function to achieve a smooth transition from the resting airfoil to the plunging motion.
This corresponds to an induced angle of attack $\alpha=21.5 \degree$, which is sufficient for the formation of unsteady leading-edge separation and dynamic-stall-like vortices.

\begin{figure}[!t]
        \centering
        \begin{subfigure}{0.49\textwidth}
			\includegraphics[width=\textwidth]{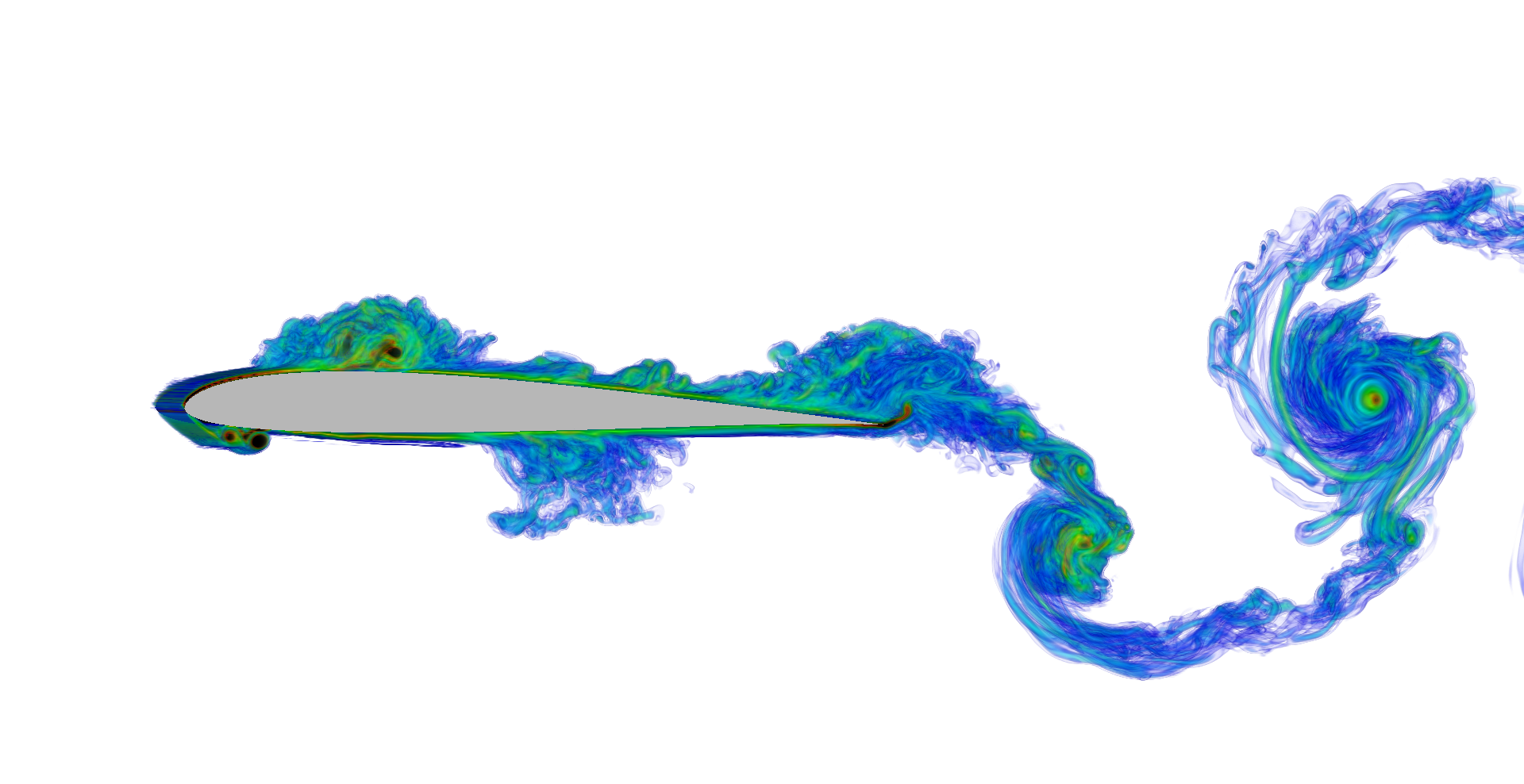}
			\caption{$\varphi=0$ }
			\label{fig:phi0}
		\end{subfigure}~
		\begin{subfigure}{0.49\textwidth}
			\includegraphics[width=\textwidth]{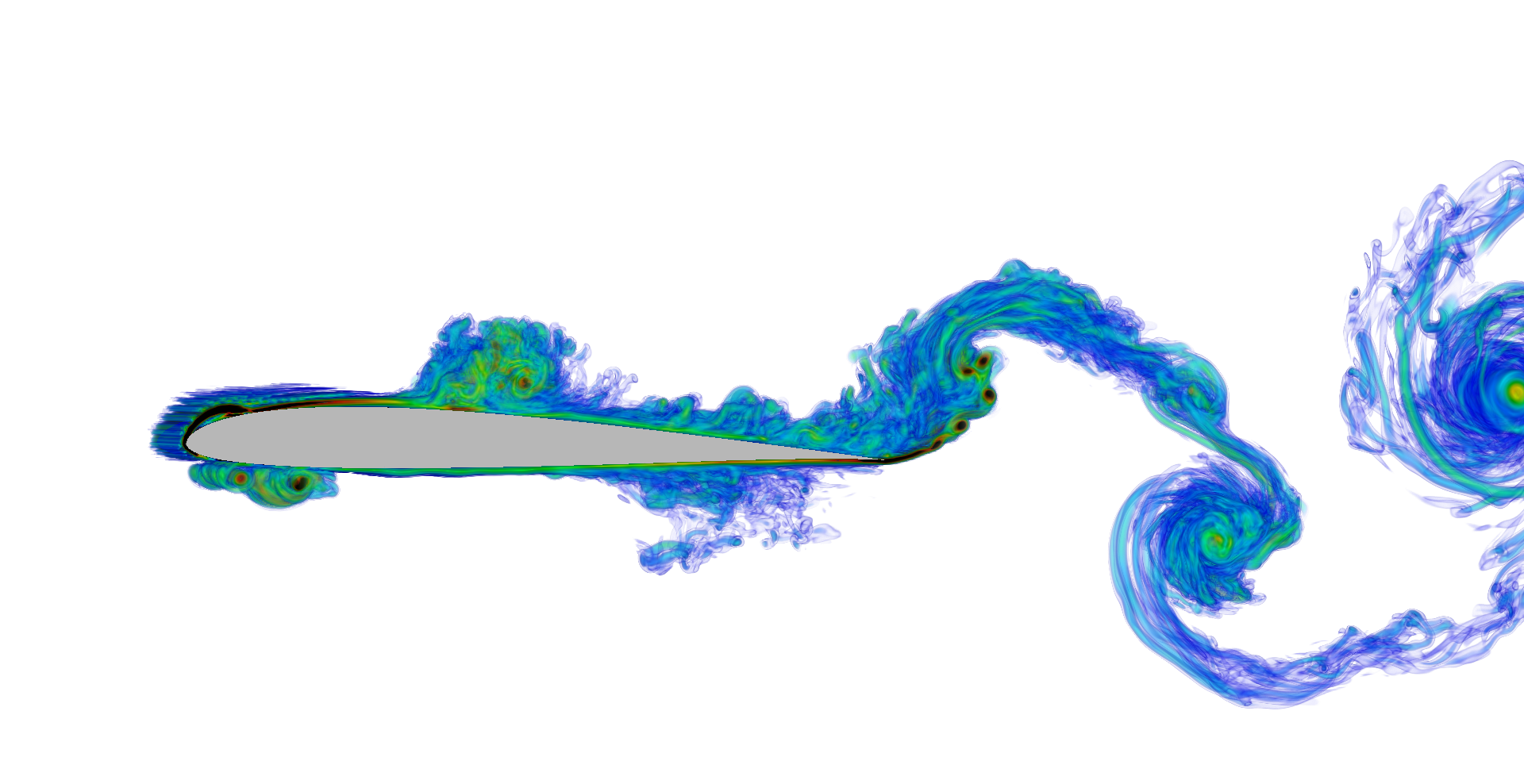}
			\caption{$\varphi=0.25$ }
			\label{fig:phi025}
		\end{subfigure}

		\begin{subfigure}{0.49\textwidth}
			\includegraphics[width=\textwidth]{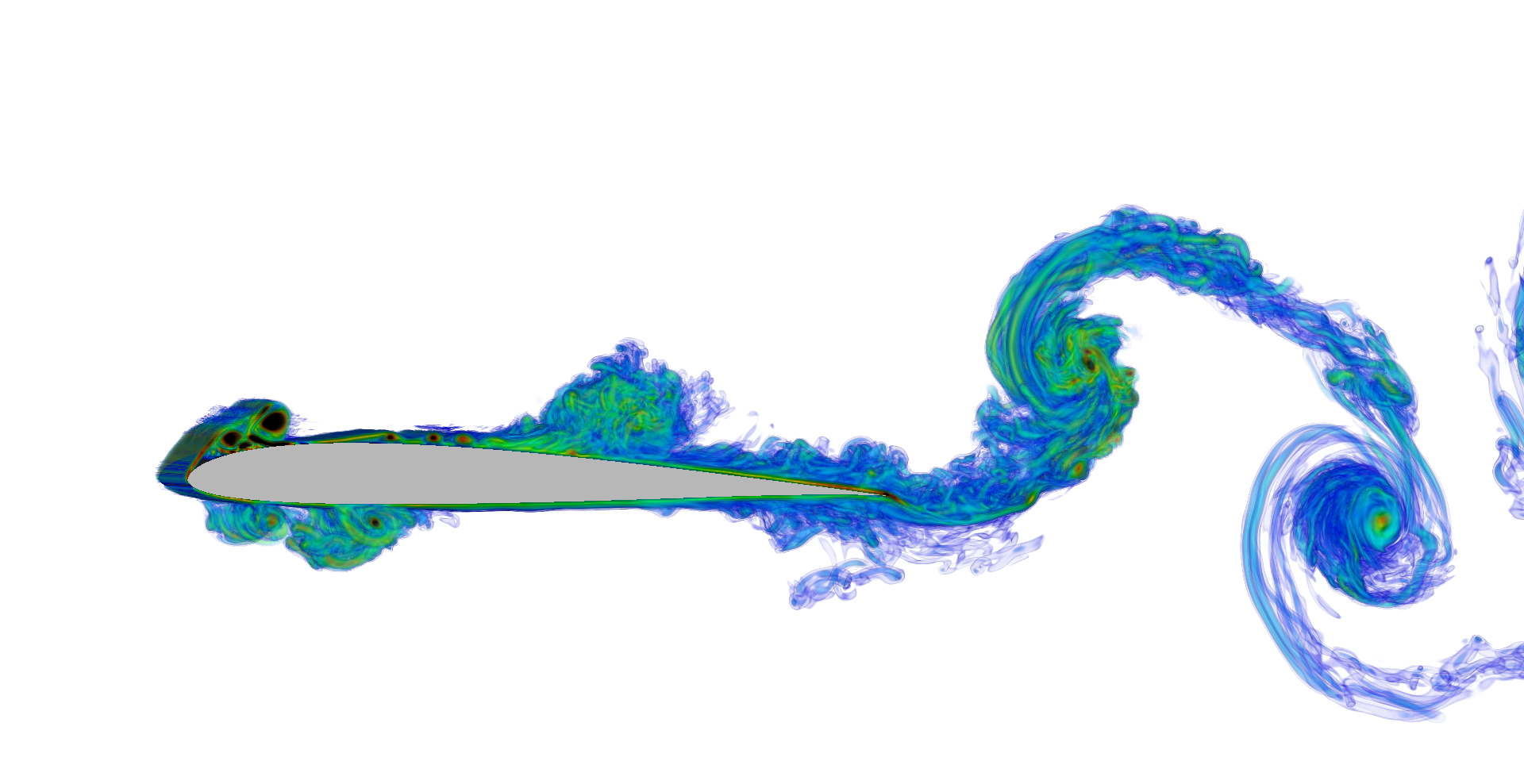}
			\caption{$\varphi=0.5$ }
			\label{fig:phi05}
		\end{subfigure}
		\begin{subfigure}{0.49\textwidth}
			\includegraphics[width=\textwidth]{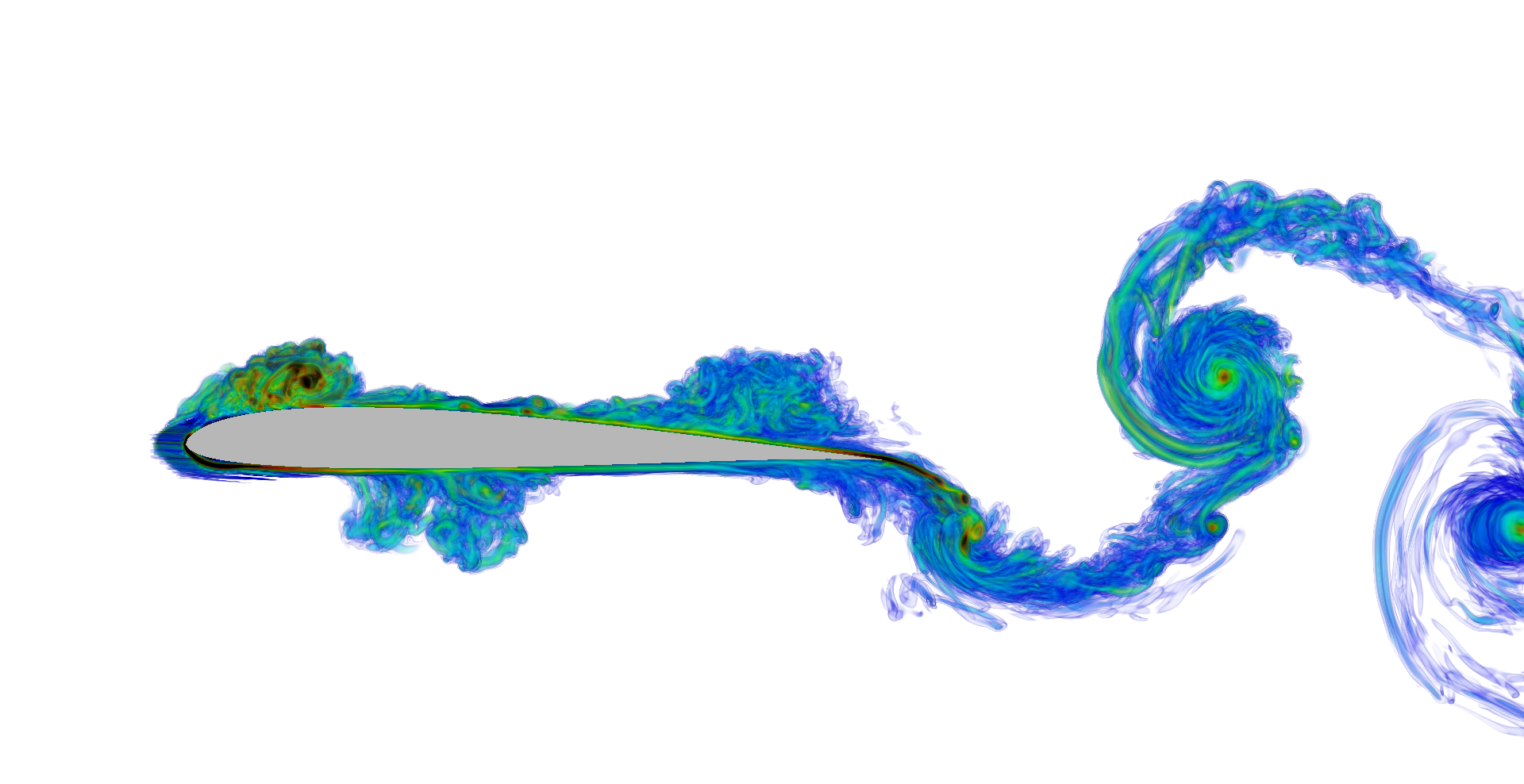}
			\caption{$\varphi=0.75$ }
			\label{fig:phi075}
		\end{subfigure}
		\caption{Volume rendering of vorticity for various phases of the plunging airfoil.}
		\label{fig:snapshots}
\end{figure}

\begin{figure}[!t]
\centering
\includegraphics[width=0.9\textwidth]{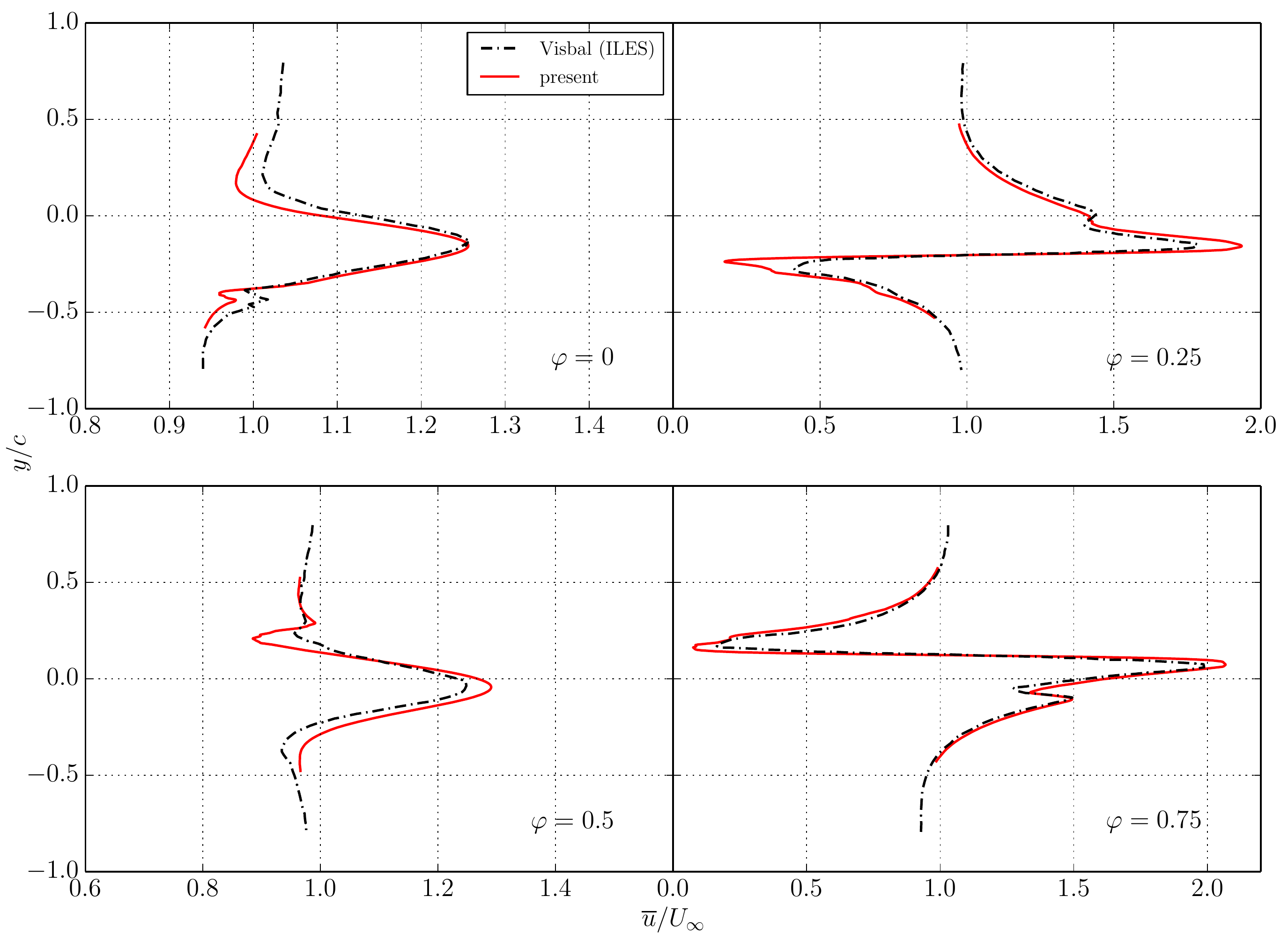}
\caption{Phase-averaged velocity profiles for $x/D=1.5$.}
\label{fig:airfoil_profiles}
\end{figure}

\begin{figure}[!t]
\centering
\includegraphics[width=0.9\textwidth]{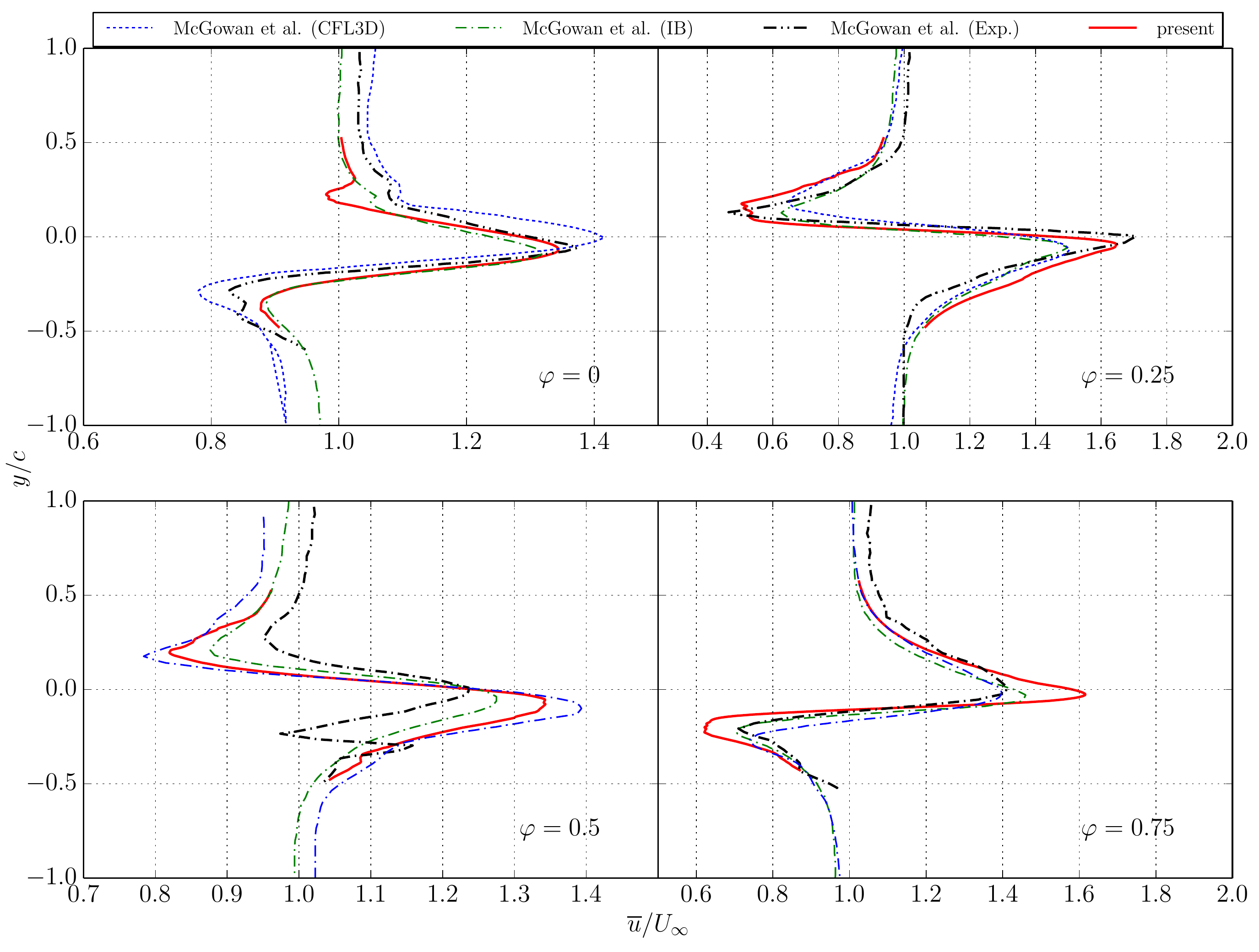}
\caption{Phase-averaged velocity profiles for $x/D=2$.}
\label{fig:airfoil_profiles2}
\end{figure}

\begin{figure}[!t]
\centering
\includegraphics[width=0.7\textwidth]{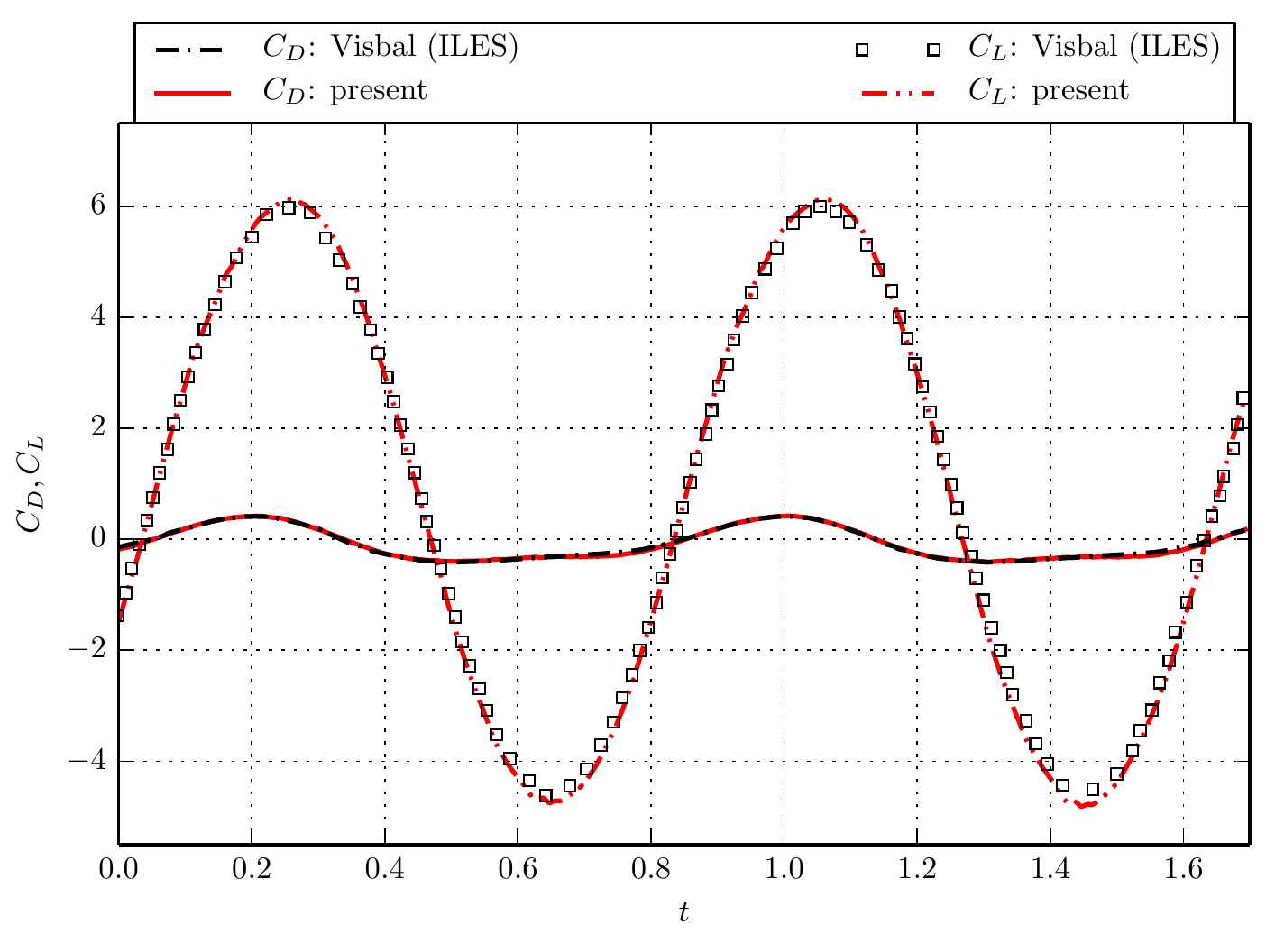}
\caption{Evolution of lift and drag coefficient over three exemplary cycles of a plunging airfoil.}
\label{fig:airfoil_forces}
\end{figure}

Capturing the main flow features occurring during the periodical motion of the airfoil, Fig.~(\ref{fig:snapshots}) shows four volume renderings of 
the instantaneous vorticity for phase angles of $\varphi=\{ 0,\frac{1}{4},\frac{1}{2},\frac{3}{4} \}$,
corresponding to maximum upward displacement, maximum downward velocity, maximum downward displacement and maximum upward velocity, respectively (see also Fig.~(\ref{fig:schematic_airfoil})).
At the beginning of a new cycle, at a position of maximum upward displacement, the boundary layer near the leading-edge appears laminar and attached to the surface (see Fig.~(\ref{fig:phi0})).
At later times, caused by the downward acceleration, an emerging flow separation can be observed at the leading-edge (see Fig.~(\ref{fig:phi025})).
At bottom dead center (BDC), the flow is fully separated at the leading-edge on the upper surface of the airfoil, causing the formation of two coherent vortices (see Fig.~(\ref{fig:phi05})). 
However, due to spanwise instabilities, these vortices break down into three dimensional, fine-scale turbulence during the subsequent upward acceleration (see Fig.~(\ref{fig:phi075})).
While diffusing and annihilating, these vortices propagate close to the airfoil surface during the following cycle. 
Note that due to the high frequency, a new pair of vortices is formed before the vortex pair of the previous cycle is shed from the trailing edge.
Identical but less pronounced is the flow structure on the lower surface of the airfoil, where the large negative motion-induced angle of attack causes 
the formation of two coherent vortex structures, which subsequently break down into turbulence. 
Similar observations were reported by \citet{Visbal2009}.\\
More quantitatively, in Fig.~({\ref{fig:airfoil_profiles}}), we compare the phase-averaged velocity profiles in the near wake  of the airfoil at $x/c=1.5$ with the study of \citet{Visbal2009}. 
In total 25 cycles were computed and the first ten were neglected in the accumulation of statistics to avoid accounting for the initial transient. 
It is apparent that the main flow features are captured for both simulations and the agreement is good.
The location of the minimum and maximum of the flow velocity agree well but differ slightly in terms of magnitude.
The present simulation exhibits larger velocity magnitudes in comparison with the more smoothed ILES simulation.
In addition, a comparison at a location further downstream at $x/c=2$ with an experimental investigation using Particel Image Velocimetry (PIV), 
an immersed  boundary method and the NASA-CFL3D code from the study of \citet{McGowan2008_AIAA}  is shown in Fig.~(\ref{fig:airfoil_profiles2}).
It is apparent that due to the complexity of the problem all data have notable discrepancies but the agreement of the present study with the experiment is reasonable.\\
Finally, we compare the evolution of the aerodynamic forces for three selected cycles with \citet{Visbal2009} in Fig.~(\ref{fig:airfoil_forces}).
The lift coefficient $C_L$ is dominated by the prescribed motion of the airfoil and has a large amplitude.
On the other hand, the amplitude of the drag coefficient is much smaller but notably there is a net mean thrust corresponding to 
a mainly negative drag coefficient.
The comparison with the ILES is excellent.

\subsection{Anguilliform Swimmer}
\label{sec:swimmer}

In the field of bio-fluidmechanics, the topic of aquatic animal propulsion mechanisms is much discussed 
among biologists, neuro-scientist as well as engineers trying to mimic these mechanisms to increase efficiency 
in technical applications (see, e.g., \citet{Ekeberg1993,Ijspeert1999, Tytell2004}).
However, due to the complex interaction between the fluid and the deformable body of the animal, fundamental questions regarding 
thrust generation and its relation to the kinematics of the swimmer's body, the efficiency and viscous effects remain controversial. 
In particular, investigations of the undulatory propulsion mechanisms of anguilliform fish (e.g.: Anguilla anguilla) started with the pioneering 
work of \citet{Gray1933}. In contrast  to carangiform fish, anguilliform swimmer generate thrust by passing a transverse wave down their body and
therefore utilizing, to a varying degree, the whole body for thrust generation and not just the tail.
The hydrodynamics of aquatic locomotion for undulatory swimmers was studied for inviscid flow in the works of \citet{Lighthill1960,Lighthill1969,Lighthill1970,Lighthill1971,Wu1971, Hess1984, Cheng1991}.

{Simple algebraic models predict a reverse von Karman vortex street for optimized swimming performance, where the wake consists of a double row of single vortices when the ratio of the swimming speed and the body wave speed is less 
than one (\citet{Lighthill1969}).}

On the experimental side, Particle Image Velocimetry (PIV) has proven itself as a valuable tool to quantify the flow field generated by the swimmer. 
The first visualization using two-dimensional PIV for freely swimming juvenile eel was reported by \citet{muller2001}.
They found a linearly increasing flow velocity from head to tail, suggesting continuous thrust generation along the body. 
In the wake, they observed a double row of double vortices with little backward momentum, which is generated by a start-stop vortex shed from the tail 
and a separate vortex produced along the body, so-called proto-vortcies, for each half tail-beat. 
They conjectured that the wake morphology is caused by a phase lag between the primary start-stop vortex and the body-generated circulation.
Subsequent studies using high-resolution PIV were conducted by \citet{Tytell2004}. 
Their results were, in general, similar to \citet{muller2001} but differences were noticed regarding the proto-vortices.
Only a negligible phase lag with low vorticity was reported, resulting in a single, combined primary start-stop vortex per half tail-beat.
The discrepancies were attributed to the lower PIV resolution and the seemly accelerating eel in \citet{muller2001} compared to the steadily swimming fish 
in \citet{Tytell2004}.
In the work of \citet{Tytell2004}, they stated the following mechanism for the generation of the wake:
A primary start-stop vortex is shed when the tail changes direction. Due to the acceleration of the tail from one side to the other, 
a low pressure region develops in the posterior part of the body, drawing fluid in the lateral direction, which is shed off the tail 
and stretches the primary vortex into an unstable shear layer which rolls up into two separate, co-rotating vortices, which they termed the secondary vortex.
Thus, the primary vortex from one half tail beat and the secondary vortex from the subsequent comprise the boundaries of each lateral jet.
Notable was the lack of any significant downstream flow for steady swimming as previously observed for carangiform fish and interpreted as thrust generation mechanism. This was further supported by the observation of only a slight upstream inclination of the lateral jets.

On the numerical front, only a few detailed studies may be found in literature (see , e.g., \citet{Carling1998,Kern2006,Borazjani2008,Borazjani2009}). 
The first two-dimensional viscous flow simulations of self-propelled anguilliform fish were reported in the work of \citet{Carling1998}.
In contrast to the experimentally observed wake morphology, these simulations indicated a single, large vortex ring wrapping around the eel, resulting in an upstream flow, where thrust is almost exclusively produced along the body and not the tail tip. 
Three dimensional simulations were recently conducted by \citet{Kern2006} using the finite volume approach of the commercial software package STAR-CD with a first-order discretization in time and second-order in space.
Apart from prescribing a reference motion of the fish, as proposed by \citet{Carling1998}, an evolutionary algorithm was employed to
obtain the body motion as a result of optimizing for burst swimming speed and efficiency.
Their results indicate that large amplitude tail undulation in combination with a straight anterior body produce most of the  thrust at the tail and are optimal kinematics in the burst swimming mode.
On the other hand, optimal kinematics for efficient swimming were obtained for an undulation of the entire body, where thrust is generated 
with half the body and not just the tail. 
For all swimming patterns, the wake morphology did not differ qualitatively and is in agreement with the experimental 
observations of \citet{Tytell2004}, exhibiting a double row of single vortex rings with lateral jets.\\
Despite the valuable contributions mentioned above, further quantitative analysis is needed for conclusive results.
The ease of data extraction, its analysis and the precise control over the body's kinematics make numerical 
experiments useful for these studies.
However, issues related to deformable meshes and the complex fluid-structure interaction are challenging for numerical solvers, thus explaining
the sparsity of these simulations in literature.
%
The KBC model on the other hand is a highly efficient approach, which allows for a simple implementation of complex, deformable bodies using Cartesian meshes 
and is therefore ideally suited for this type of problem. 
{The moving wall boundary condition of \citet{Dorschner2015} allows for non-body fitted meshed, which significantly reduces and simplifies the process of 
grid generation. This, combined with the superior stability of KBC models, the intrinsic parallelizability and efficiency of LBM make these the best available tools
for DNS investigations of such complex flows. }
Therefore, as a first step to access the predictive capabilities of the proposed scheme, we conduct 
a simulation of a self-propelled anguilliform swimmer analoguous to \citet{Kern2006}.\\
\begin{figure}[!t]
\centering
\includegraphics[width=0.9\textwidth]{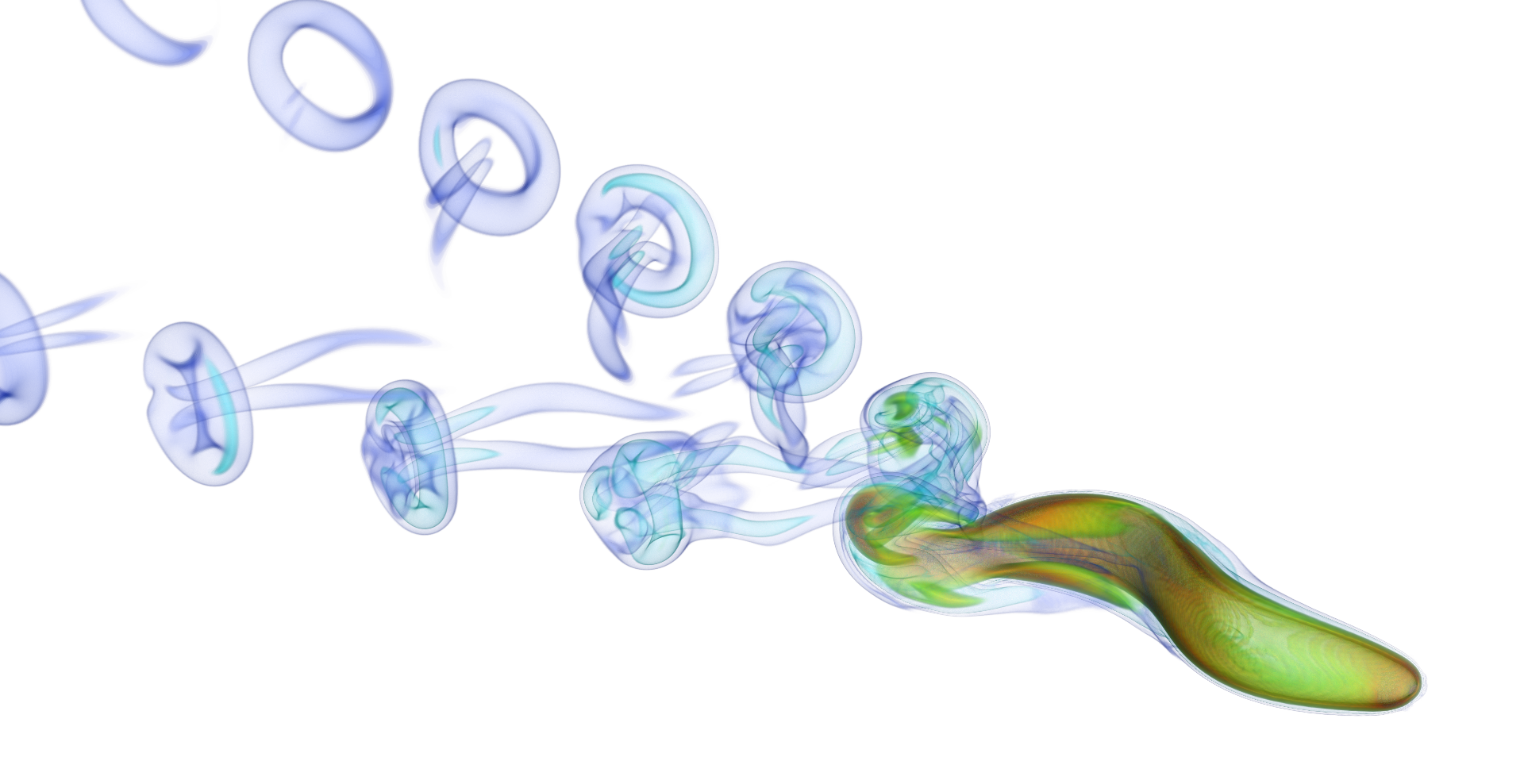}
\caption{Volume rendering of vorticity of the swimmers wake, showing the typical double row vortex street.}
\label{fig:fish_volRendering}
\end{figure}
%
%
%
%
%
Geometrically, the body of the anguilliform swimmer with length $L$ is modeled by spatially varying ellipsoidal cross sections as in \citet{Kern2006} 
for which the half axes $w(s)$ and $h(s)$ are defined as analytical functions of the arc-length $s$ as 
\begin{equation}
w(s)=
\begin{cases}
      \sqrt{2w_hs-s^2} \qquad \qquad  & 0 \leq s<s_b,\\ 
		w_h-(w_h-w_t) \left( \frac{s-s_b}{s_t-s_b}\right)^2 \qquad  \qquad  &s_b \leq s < s_t,\\
      w_t \frac{L-s}{L-s_t} & s_t \leq s \leq L,
\end{cases}
\end{equation}
where $w_h=s_b=0.04L$, $s_t=0.95L$ and $w_t=0.01L$.
For the height $h(s)$ an elliptical curve is prescribed as
\begin{equation}
	h(s)= b \sqrt{1- \left( \frac{s-a}{a} \right)^2},
\end{equation}
where the half axes are given as $a=0.51L$ and $b=0.08L$.
As proposed by \citet{Carling1998}, the swimmer undergoes a prescribed time-dependent lateral deformation of the body's center-line as
\begin{equation}
	y(s,t) = 0.125 \beta  L \frac{s/L+0.03125}{1.03125}\sin{ \left (2\pi \left( \frac{s}{L}- \frac{t}{T} \right) \right)},
\end{equation}
where $T$ is the undulation period and $\beta$ is an initial ramping function to assure a smooth transition 
from the initially resting body according to 
\begin{equation}
	\beta=
	\begin{cases}
		\frac{1-cos(\pi t/T )}{2} \qquad & 0 \leq t \leq T, \\
		1 \qquad & t >T.
	\end{cases}
	\label{eq:fish_ramping}
\end{equation}
The feedback from the fluid onto the body is prescribed by Newton's equations as 
\begin{align}
	m_s \ddot{{\bm x}_s} &= \bm F \\
	\frac{ {\rm d} \bm I_s \bm \omega_s}{ {\rm d} t} &= \bm M,
\end{align}
where $m_s$, $\bm I_s$, $\bm x_s$ are the mass, the inertia tensor and the center of mass, respectively. 
The force and torque acting from the fluid onto the object are denoted by $\bm F$ and $\bm M$, respectively.
As in the work of \citet{Kern2006}, the complexity of the system is reduced by only considering the torque corresponding to 
the yaw.
In the framework of the LB simulation, the geometrical model is represented as a triangulated surface composed of elliptical disks along the center-line.
The center-line was decomposed into $150$ segments, which has proven to be sufficient.
This allows for a straightforward computation of all geometrical properties including the time derivative of the inertia tensor.
The velocity at the intersection points, needed for the boundary conditions, was computed based on a finite-difference scheme and a barycentric interpolation for the given time-dependent deformation.
As above, the equations of motion are solved using an Euler integration scheme. Higher-order integration schemes have been tested but 
marginal differences were observed due to the relatively small time step used. 
Further, no smoothing or low-pass filtering of the hydrodynamic forces were applied, unlike in \citet{Kern2006}.

The simulations are carried out with a uniform grid using a domain of $[ 8L \times 4L \times L ]$, the swimmer is resolved by $L=200$ lattice points 
and the undulation period is taken to be $T=10^4$ lattice time steps.
Same as in the reference, the Reynolds number is taken as ${\rm Re}=(L^2/T)/\nu = 7142$.
\begin{figure}[!t]
\centering
\includegraphics[width=0.7\textwidth]{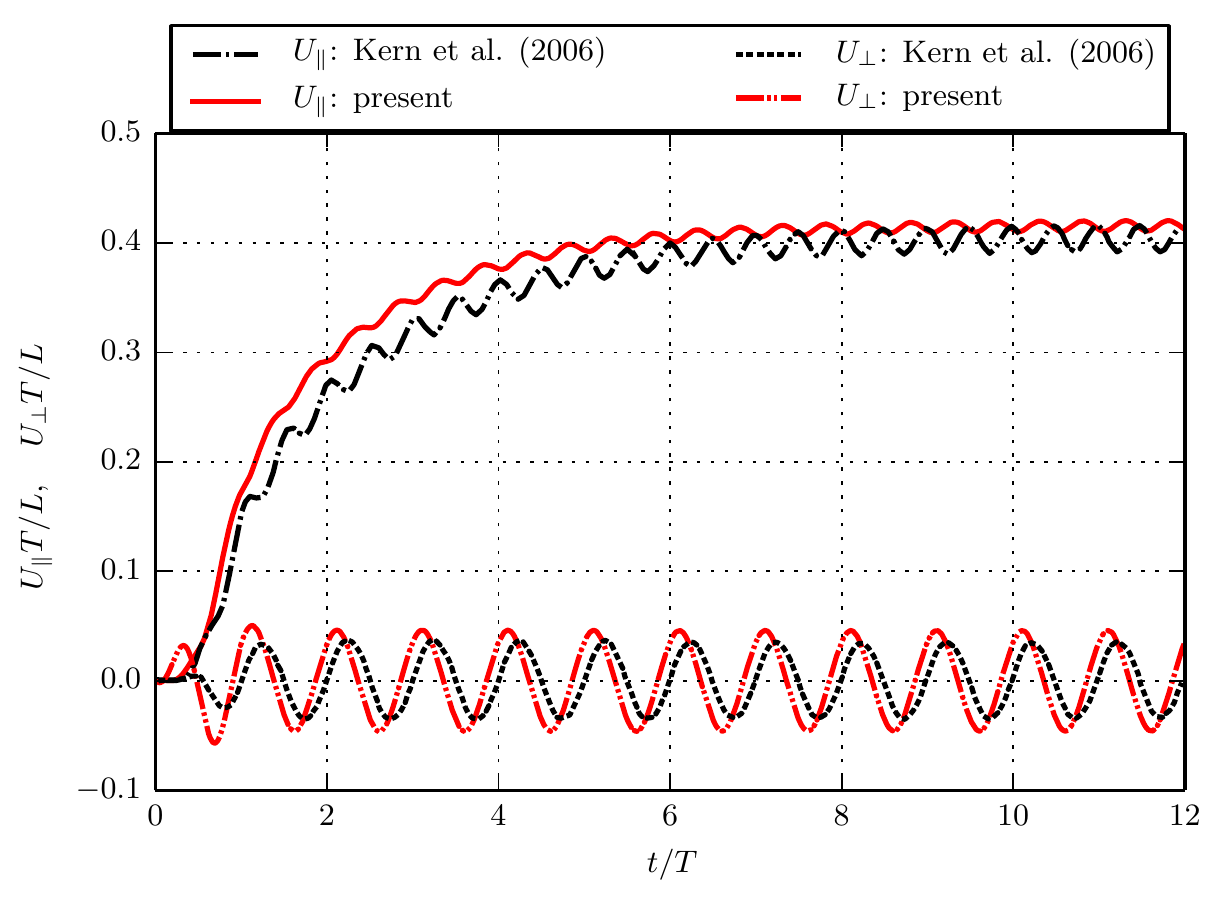}
\caption{Temporal evolution of the forward velocity $U_\parallel$ and lateral velocity $U_\bot$.}
\label{fig:fish_vel}
\end{figure}

A volume rendering of vorticity shows the wake of the swimmer in Fig.~(\ref{fig:fish_volRendering}).
In agreement with \citet{Kern2006, Tytell2004, tytell2004a}, the wake consists of
a double row of single vortices. Furthermore, we confirm the measurements from \citet{Tytell2004} and observe a primary vortex being shed from the tail when it
changes direction. The fluid drawn in lateral direction by the accelerating tail and the development of two separate co-rotating vortices as a result of an unstable shear layer roll-up can also be seen in Fig.~(\ref{fig:fish_volRendering}).

Quantitatively, we compare the evolution of the forward and the lateral velocity of the swimmer in Fig.~(\ref{fig:fish_vel}).
Overall, the agreement is good.
Minor, discrepancies are observed during acceleration but are attributed to different ramping functions used here, Eq.~(\ref{eq:fish_ramping}), and in \citet{Kern2006}.
The differences in the asymptotic forward and lateral velocity are within the range of expectation due to the different evolution algorithms.  
While \citet{Kern2006} report an asymptotic forward velocity of $\bar{U}_{\parallel}=0.4$ with an amplitude of $0.01$, the present simulation yields
$\bar{U}_{\parallel}=0.415$ with an amplitude of $0.005$.
The lateral velocity $U_\bot$ has a zero mean for both the reference and the present simulation. An amplitude of $0.046$ is measured in the present simulation 
whereas \citet{Kern2006} measure $0.03$.
All simulations were checked for grid independence. 

\section{Conclusion}
\label{sec:conclusion}
In this contribution we have presented a thorough study of the entropic multl-relxation time model in combination with the Grad boundary condition 
for moving and deformable objects in three dimensions. 
After validation using the classical benchmark of a sedimenting sphere under gravity, the accuracy and robustness of the proposed method was accessed in the 
simulation of a plunging airfoil in the transitional regime. The comparison with literature revealed excellent agreement in all quantities including 
 phase-averaged velocity profiles in the near wake region as well as the evolution of aerodynamic loads, important for realistic two-way coupled simulations.
Finally, for deformable objects, the simulation of an anguilliform swimmer was considered and validated by comparison to the numerical investigations of \citet{Kern2006}.
These simulations together with the previous studies (\citet{Bosch2015,Chikatamarla2013, Dorschner2016}) establish the predictive capabilities of KBC models for complex, moving and deforming objects, where a simple grid-convergence study is sufficient to assert their validity.
The above simulations of complex flows, when viewed together with the efficiency of KBC models and the ease of implementing boundary conditions, demonstrate that the present method may be a viable alternative for engineering-related fluid dynamics.


\section*{Acknowledgments}
This work was supported by the European Research Council (ERC) Advanced Grant
No. 291094-ELBM and the ETH-32-14-2 grant. The computational resources at the Swiss
National Super Computing Center CSCS were provided under the grant s492 and s630.

\bibliographystyle{plainnat}
\bibliography{refs}

\end{document}